# COSTS AND DIFFICULTIES OF LARGE-SCALE 'MESSAGING', AND THE NEED FOR INTERNATIONAL DEBATE ON POTENTIAL RISKS

John Billingham, SETI Institute, Mountain View, CA, USA
James Benford, Microwave Sciences, Lafayette, CA, 94549 USA
jbenford@earthlink.net

Abstract

We advocate international consultations on societal and technical issues to address the risk problem, and a moratorium on future METI transmissions until such issues are resolved. Instead, we recommend continuing to conduct SETI by listening, with no innate risk, while using powerful new search systems to give a better total probability of detection of beacons and messages than METI for the same cost, and with no need for a long obligatory wait for a response. Realistically, beacons are costly. In light of recent work on the economics of contact by radio, we offer alternatives to the current standard of SETI searches. Historical leakage from Earth has been undetectable as messages for credible receiver systems. Transmissions ('messages') to date are faint and very unlikely to be detected, even by very nearby stars. Future space microwave and laser power systems will likely be more visible.

key words: METI, SETI, beacons, microwave

## 1.0 The METI Debate

There are two basic approaches for communicating between the stars: receiving and transmitting (Hawking 2010). The first uses telescopes to attempt to detect electromagnetic signals transmitted by an extraterrestrial civilization. The majority of searches carried out to date, numbering about 100, fall into this category, and form the basis of the long-term program of the SETI Institute, as well as other organizations. The second is for us to transmit messages from Earth to promising targets in the Galaxy, in the hope that they may be detected by extraterrestrial intelligent species, who in turn might respond to us. This is Messaging to Extraterrestrial Intelligence, METI. It is also known as Active SETI. It has been proposed or used occasionally in the past, but is now proceeding apace under the aegis of Alexander Zaitsev at the Evpatoria telescope in the Ukraine.

Stephen Hawking, in a show about Aliens on the Discovery Channel, thinks that it might be unwise to transmit. He reminded us that if our METI signal were to be detected by a civilization considerably older than we are, it might have unlimited power, with the ability, for example, to harness the entire output of its star, warp space, create worm holes, or travel huge distances in the blink of an eye. Arthur Clarke put it this way: "Any sufficiently advanced technology would be indistinguishable from magic". Statistically, it is likely that such a species would be much older than us, because we are essentially the youngest communicating civilization in the Galaxy. It is impossible for us to know today what is the societal, cultural, political and ethical make up of such a society. As many have pointed out, they could be a benign and benevolent race, whose main interest in the comparatively primitive beings of Earth, namely us, is to foster long-distance intellectual exchanges with kindness and munificence.

On the other hand, Hawking is asking if such contact could pose any risk to us. He used an old analogy, that past contacts between the peoples of Earth has frequently resulted in disastrous results for the less technologically advanced races, and quoted the unhappy fate of may of the indigenous tribes of the Caribbean in the centuries after Columbus first landed there. Similar devastating consequences have resulted in many places on Earth. If the extraterrestrial species encountered because of establishing future long-term METI endeavors of great magnitude and cost were somehow to fit this bill, what might become of us? We cannot be specific about the possible risks involved, but note Hawking's concerns that the other civilization might have powers that would be beyond our own understanding, and interstellar mores and policies that are unknown. Some have argued that many of our current microwave transmissions leak out into the Galaxy anyway, and will be detectable at increasing distances with the passage of time, and with the rise in transmitter powers by one to two orders of magnitude within a couple of decades. This just adds to any risk from METI itself.

*1.1 The Need for International Consultations*

*The time has come to address these issues on an international scale, by establishing international symposia on all aspects of transmitting from Earth to extraterrestrial intelligence.* The goal would be to reach a consensus on what to do about METI, and future leakage from Earth. Should decisions just be left in the hands of individuals or small groups, or national organizations? Should there be informal or formal agreements between nations to take action to resolve risk issues? When should agreements be translated into action? *We should begin the international symposia soon.*

Such international gatherings, with experts and non-experts from all disciplines, would cover a wide range of topics related to METI and SETI. Risk analysis would be a driver, but there is a long list of other topics:

- <u>Technical:</u> science and engineering of beacons; detection probabilities; standard "passive" SETI as it exists now, and in the future, as compared to METI.
- <u>Sociological:</u> study of historical analogs, sociological issues, national and international law, education and the media.

*1.2 Historical Examples of Such Methods*

<u>The UN:</u> In 2000 the International Academy of Astronautics sent a proposal to the UN Committee on the Peaceful Uses of Outer Space entitled "Declaration of Principles for Activities Following the Detection of Extraterrestrial Intelligence", also known as the *First Protocol* (Billingham and Heyns 1999). The proposal was received without objection. Principle 8 reads, in part "No response to a signal or other evidence of extraterrestrial intelligence should be sent until appropriate international consultations have taken place". No one seems opposed to having international consultations about transmitting *after* we detect them by standard SETI. Assuming this to be the case, it is surely even more important to have the consultations about transmitting *before* we detect them when we don't even have their signal in hand.

The USA: Because no one can say there is zero risk from biohazards, NASA has for a long time had a Planetary Protection Office, which has rules and regulations about designing and carrying out planetary sample return missions, so that the Earth is not exposed to extraterrestrial biohazards (Conley 2010). This Office works in close conjunction with the Committee on Space Research (COSPAR) of the International Council for Science, so that now the 'protection of Earth' policy is international. For the Apollo 11 mission, NASA elected to quarantine the crew after they returned to Earth (Primary Mission Accomplished: 1969).

*1.3 Moving Forward on METI*

There are many possible organizations for organizing such Symposia: International Astronomical Societies, the American Association for the Advancement of Science, the International Institute of Space Law, or the Royal Society and, as before, the UN Committee on the Peaceful Uses of Outer Space is an obvious candidate for participating in the debate.

Note that Hawking did not say that there was a risk in just listening - the standard approach to SETI today. In fact, there is no innate risk in just listening, except culture shock. That is, unless you include science fiction stories about the alien messages which include instructions on how to assemble pathogenic organisms, or even replicas of evil aliens themselves. But we are not obliged to respond to ETI. Regardless of how determined Zaitsev and others may be to undertake METI because they feel that conventional SETI is not working, or for other reasons, it should be recognized that we have still only scratched the surface of the multi-dimensional search space of current SETI. As of now, we have covered a volume in 1.6 cups of water, as compared with the volume of all the world's oceans (Tarter, 2010, section 3). Let us continue until we have at least excavated more of this huge search space. Let us have the international symposia to try to resolve the METI issues, and soon.

Note that no one can say that there is zero risk to transmitting. So, as the authors of this paper, we agree with Hawking that it may be unwise to transmit, at least unless and until the international debate generates consensus findings on the issue.

We advocate a moratorium on METI until an international consensus is reached.

**2.0 Building and Searching for Beacons**

*2.1 Beacon Cost-Why Economics Matters*

Reaching out on large scales > 1000 ly demands far more aperture, power and expense than the small-scale transmissions done to date using radio telescopes. Economics will matter crucially (Benford J. et al. 2010). The difference between peak and average power costs has an impact on beacon design. There are significantly lower costs for pulsed sources for fundamental physics reasons. The physics for the cost reduction is that the electrical breakdown threshold is much higher for short pulses, so much more energy can be stored in small volumes. For pulsed peak power systems, costs are in the range of 0.1-0.01$/W. Then a GW power system costs 10-100 M$. High average power systems have costs driven by continuous power-handling equipment and cooling for losses. They cost ~3$/W, so a GW unit costs ~3B$.

If the nearest ET civilization is 1,000 ly from us, the cost of building a beacon to announce ourselves to them would be 6.8 B$ for continuous and 1.2 B$ for short-pulse beacons. The scaling is shown in Figure 1.

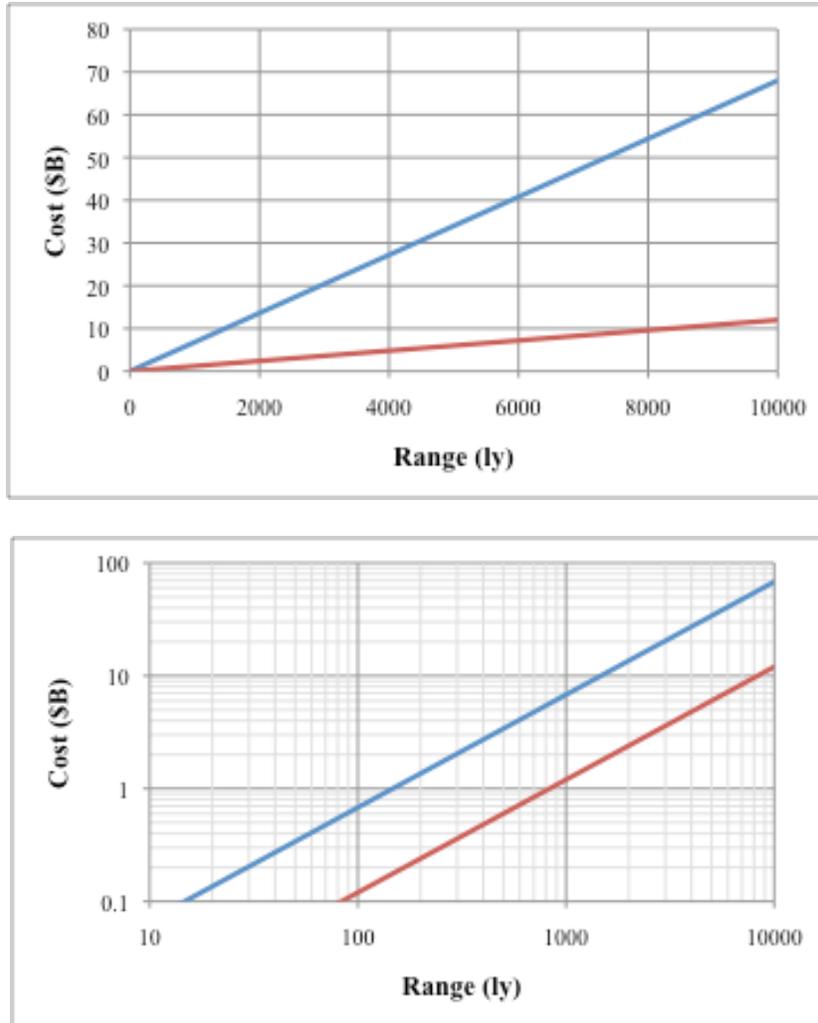

Figure 1 Cost scaling of cost-optimized beacons on linear and log-log scales, based on current Earth costs for antennas and microwave power.

Contrast that with the Square Kilometer Array (SKA), which is currently estimated to cost $470,000 per dish/electronics module. For 3,000 dishes, the cost is 1.4 B$. Data analysis and ancillary equipment bring the total to ~2 B$. The range for the SKA to see pulsars will cover the whole galaxy. Pulsars look a lot like beacons, so SKA can search for beacons across the whole galaxy for about the cost of a 1,000 light year range beacon.

*2.2 Searching for Beacons*

Advocates have for some time made a case for the value of searching the 'dynamic sky' for transients that have been observed but not frequently enough to be understood. SETI signals will in fact be more like transients and less like the fairly steady beacons SETI has traditionally assumed. Search methods directed at such short transient signals will be more likely to see cost-optimized beacons. Past SETI surveys couldn't detect such pulsed signals well, because they integrate over time and had limited frequency coverage, usually just 'waterhole' frequencies. The cardinal rule of observational astronomy is that a signal has to be seen to repeat to be believed. But that condition on the signal is also a requirement on the listener. We must be persistent and listen for a significant time.

The most efficient approach is to look steadily for transients in the galactic plane for a long time over a large frequency range and high frequency response. The key point is that it's a staring array: each dish has its assigned piece of the plane to survey. The galactic plane is the place to look for short transients, as the recent Astropulse data analysis has shown. In addition, the Life Plane strategy makes an argument that the increasingly accepted model of Medvedev & Melott for the observed periodicity of extinction events in marine life means that life, hence civilizations, and hence SETI beacons will be more tightly clustered to the plane than the stars themselves. (Benford, G. et al., 2010).

Therefore, a small array of Australian SKA Prototype (ASKAP) dishes could survey the plane continuously. This can be done inexpensively by using SKA technologies, taking advantage of economies of scale. Estimated cost is less than 1% of the SKA cost, ~20 M$.

**3.0 Analysis of Zaitsev Messages**

Can recent deliberate transmissions from Earth be detected at the distances of nearby stars? What is the reality of claims that the fairly low power messages sent to date are 'practically detectable everywhere in the Milky Way' (Zaitsev 2008)? If not quantified, such qualitative statements are not very useful in a quantitative science. We will see that the messages are faint and very unlikely to be detected, even by very nearby stars.

We take as an example the Cosmic Call 1 message to the stars Zaitsev (2006) radiated in 1999 from the RT-70, Evpatoria, in Crimea, Ukraine, a 70-m dish with transmitter power up to 150 kW at frequencies about 5 GHz. The content ranges from simple digital signals to music. Can civilizations in the stars hear them? The stars targeted range between 32 ly and 70 ly, so the signals will be weak when they arrive. The question then becomes: how big an antenna and how sensitive receiver electronics need be to detect them? And will it be received as a message or a long energy pulse?

At what ranges will high and low data rate messages be detectable? At what ranges will only message energy be detectable? For the receiver, at what ranges would present Earth technology see the messages from Earth? At what ranges would the future SKA see messages from Earth?

*3.1 Signal Strength*

The power density S at range R is determined by W, the effective isotropic radiated power (EIRP), the product of radiated peak power P and aperture gain G,

$$W = PG$$
$$S = \frac{W}{4\pi R^2} \quad (1)$$

and gain is given by area and wavelength:

$$G = \frac{4\pi\varepsilon A}{\lambda^2} = 5.18\left[\frac{D_t}{\lambda}\right]^2 = 7 \times 10^6 \quad (2)$$

where $\varepsilon$ is aperture efficiency and the result is for dish construction. Then $W=10^{12}$ watts (~1/10 that of Arecibo), and the power density at light year distances will be

$$S = \frac{W}{4\pi R^2} = \frac{9.3 \times 10^{-22} \text{W}/\text{m}^2}{R^2(\text{ly})} \quad (3)$$

At 70 ly, the intensity will be $\sim 2 \times 10^{-25}$ W/m$^2$.

*3.2 Can the Transmissions Be Detected as Messages?*

The detectability of the messages depends greatly on the bandwidth of the transmission. Low data rates, though being able to show that the signal is artificial, also mean little information. High data rates require high bandwidth, and suffer more competition from noise, which is roughly proportional to bandwidth.

Zaitsev's 1999 Cosmic Call was sent at two different bit rates. First, the message was sent 3 times at 100 bits/sec, then at 2000 bits/sec. The digital information is transmitted using frequency modulation with a deviation of 48 kHz, the shift of +24 kHz corresponding to the symbol "1", the -24 kHz shift corresponding to the symbol "0." This is called *Binary Frequency-Shift Keying* (BFSK) and is a widely used, if simple, form of modulating digital information on Earth. For the 100-bit/sec messages, the range is substantially larger than for the 2 kHz rate. This is because at the low bit rate the fraction of signal energy concentrated around the ±24 kHz frequencies increases, appearing as two spikes, and the energy between the two frequencies drops. Extra-terrestrial intelligences (ETIs) could use a detector with two correlators, where each correlates one of the FM frequencies. The correlator with the highest 'score' decides if the bit was a 1 or 0. A detector of this type would integrate all the energy during the bit period, and would have a noise bandwidth similar to the bit rate of 100 Hz. This may be what Zaitsev has assumed, i.e. noise bandwidth=bit rate. But to achieve an acceptable error rate probably S/N~10 would be required.

Then, from *Project Cyclops* (Oliver and Billingham, 1996, pg. 56) the 'reference range' for a given signal-to-noise ratio (S/N) is

$$R = \sqrt{\frac{WA_r}{4\pi k T_{sys} B_t}} \frac{1}{\sqrt{S/N}} \quad (4)$$

where $B_r$ is the receiving bandwidth, receiving antenna area $A_r$ and system temperature $T_{sys}$.

*Can Earth-like radio telescopes detect Earth's radio telescopes?* First, we evaluate the ability of Zaitsev's RT-70 to detect itself, assuming ETI has the same level of capability as ourselves. Assume $T_{sys}$=20K (probably too optimistic), 100 Hz bandwidth.

$$R_{70}^{70} = \frac{9.8 ly}{\sqrt{S/N}} \quad (5)$$

For a robust S/N of 10, this is 3 ly, less than the distance to the even the nearest star targeted-or in fact any star-so the RT-70 messages would not be detected by RT-70.

*Can an ETI SKA detect Earth Radio Telescopes?* Zaitsev's assumption is that Extra-Terrestrials have SKA-like systems with quality factors that vary from $A_r/T_{sys}$ ~50,000 m² /K to 20,000 m² /K. The *Preliminary Specification of the SKA (2007)*, of 2007 gives values <10,000 m² /K for mid-range microwave frequencies.) We will take the SKA to have $A_r/T_{s\ 2ys}$=5,400 m² /K, the present experimental value of the ASKAP (DeBoer 2010). It may be increased to 8,400 m² /K. with larger dishes.

Evaluating eq. 4 for the case of ETI SKA with a signal of 100 Hz bandwidth,

$$R_{70}^{SKA} = \frac{59 ly}{\sqrt{S/N}} \quad (6)$$

For S/N=10, R=19 ly, which is not in the range of the stars targeted. For the closest star targeted, at 32 ly, this gives a S/N of 3.4 for a SKA-scale receiver. So using a low data rate and BFSK, range can be extended if large radio telescopes are assumed. Of course, one can postulate arrays larger than SKA and get larger ranges.

Note also that the following must all occur for ETI to detect this low-bit-rate signal:

1) Their system must stare in the very small part of the sky where our sun is, i.e., they must be interested in our system. (To get high sensitivity, as we have seen, the antenna area must be large, so the targeting angle is very small.) This could be because they've detected our out-of-equilibrium atmosphere, thus possible life here. This has been true for billions of years. Or they could have detected our leakage radiation. Section 1 shows this to be difficult at ranges 30-70 ly.
2) They would have to guess the bit rate of the message. Processing the stored signal with successive assumed rates and seeing which gives the best signal could do this.
3) They would have to deduce that we're using binary frequency-shift keying (BFSK) instead of another of our many modulation methods, so must analyze the received signal against a list of such stratagems.

While all the above could occur, they are by no means certain.

*3.3 Can the Energy of the Transmissions be Detected?*

The above is an analysis of *messaging*. Some Cosmic Call messages use a long duration, up to 4 hours, which could allow ETI integration time to improve the chances of detection. But this is at the expense of destroying the information content. The *energy* of the "1"s and "0"s would be accumulated over time to increase the effective S/N. But this comes at the expense of losing the information content. It will increase the range of detection. Spectral analysis would show two peaks above the noise (carrier ±24 kHz) with some spectral spreading that shows modulation. ETI might conclude from this that BFSK was in use, and maybe even estimate an approximate bit rate. But unless they are within range for receiving the individual bits, as quantified above, the S/N will be too low to extract the bits.

To quantify the effect, note that the S/N is increased by integrating over a time t in the receiving bandwidth channel $B_r$, by a factor $(B_r t)^{1/2}$, giving a signal-to-noise ratio (Kraus 1986, Benford, J. et al. 2010):

$$\frac{S}{N} = \frac{P_r}{P_n}\sqrt{B_r t} = \frac{WA_r}{4\pi R^2 k_B T_{sys} B_t}\sqrt{B_r t} \tag{7}$$

for $B_r < B_t$. If $B_r > B_t$, this becomes

$$\frac{S}{N} = \frac{WA_r}{4\pi R^2 k_B T_{sys} B_t}\sqrt{\frac{t}{B_r}} \tag{8}$$

The range then varies as $R \sim (B_r t)^{1/4}$ for $B_r < B_t$. If $B_r > B_t$, $R \sim (t/B_r)^{1/4}$. For low data rate 100 Hz channels (both transmitter and receiver) and 4 hours duration, the factor $(B_r t)^{1/4}$ is 35, increasing range by this large factor.

*Can Earth-like radio telescopes detect Earth's radio telescopes energy by integration?* Eq. 5 becomes, for substantial integration:

$$R_{70}^{70} = \frac{343 ly}{\sqrt{S/N}} \tag{9}$$

For Earth radio telescopes such as RT-70, a 70-m dish, transmitting energy to an identical ETI radio telescope with S/N=10, the range for detection is R=108 ly. So, the RT-70 used for the Cosmic Call messages could detect its own messages as packets of energy at a range of 108 ly

*Can an ETI SKA detect earth radio telescope energy by integration?* For a SKA-scale receiver with S/N=10, range is about 6 times greater, R=648 ly.

*3.4 Conclusions and Recommendations*

To summarize the detectable ranges for various cases considered here, we compare the ranges for message content and message energy for S/N=10 in the Table. For the energy case, we assume constant reception for 4 hours. These ranges are for the low (100 bit/sec) data rates. For the receiver, we show what present Earth technology would see and what the future SKA concept would see.

**Range for Detection of Cosmic Call 1 Message**

| message rate | receiving radio telescope technology | range for message content | range for message energy |
|---|---|---|---|
| low data rate (100 bit/sec) | Earth radio telescopes | 3 light years | 108 light years |
| | SKA-scale radio telescope | 19 light years | 648 light years |

The content of Zaitsev's messages will not be recoverable as messages by ETI if their radio telescopes are comparable to ours. To be observable, the receiving area must be greater than the SKA we're contemplating building, and then only a low data rates.

Extending the messages by repeating, so they last hours, allows ETI to integrate the signal, and detect its presence at ranges of 100's of light years. But that obliterates the message content, producing a recognizable pulse of energy. That could be taken by ETI as an undifferentiated energy source that could be artificial. But it cannot be characterized as a message.

Moreover, to accomplish such a reduced objective, signal-processing theory suggests that it would be more efficient to transmit a simple sine wave.

All other messages are weaker in EIRP than Cosmic Call 1, in many cases much less powerful, so are less observable.

For higher data rates, bandwidths must expand and range drops. Cosmic Call 1 also contained a higher data rate transmission of the same message for 5 minutes at 2000 bits/sec. The bandwidth was about 2 kHz, so range is reduced (eq. 4) by a factor of $(100/2000)^{1/2}=0.22$. So, for high content rate, the message will not be detectable at the target stars, but the energy will be detectable for SKA-scale receivers.

To make it easier for the scientific community to ascertain the range and other properties of broadcasts from Earth, we suggest that all those radiating provide clear documentation.

- All such radiations, past and future, should be described completely enough to ascertain their delectability as a function of assumed ETI technologies.
- Such descriptions should be required to meet peer-reviewed publication standards, *which past radiation events have not done*.
- This information should include both transmitter parameters (power, frequency, aperture,

bandwidth, frequency stability) and 'message' parameters (bit rate, keying method, error correction coding, number of message repeats, etc.).
- If they wish, radiators could also provide what they assume might be receiver parameters such as antenna area and receiving system temperature.
- *There should be an on-line database containing such descriptions in a standard uniform format.*
- A top-level summary of the radio and laser signature of Earth should be part of that database.

**4.0 Eavesdropping on Earth**

Can leakage radiation from Earth be detected at the distances of nearby stars? The detailed study by Sullivan *et al.* (1978) occurred at what was the approximate peak of Earth's signature in the microwave. He found that the range for our 1980 technology to detect our own signature was a few light years. To quantify this using the relations in the Zaitsev message section of this paper, note that the earth EIRP in the UHF video bands is ~5 MW. From eq. 3, for $W = 5 \times 10^6$ watts at 10 ly, $S \sim [5 \times 10^{-27} \text{ W/m}^2]/R^2$ (ly). The video carriers have bandwidths of ~0.1 Hz, so the flux per hertz is $S/\Delta f \sim 5 \times 10^{-26}/R^2$ (ly) in W/m²-Hz. Parkes in Australia is a typical large radio telescope and its sensitivity is $2 \times 10^{-25}$ W/m²-Hz. So, even at Alpha Centauri (R = 4.25 ly), Parkes could not see video leakage from Earth.

Picking up signals from commercial radio and television broadcasts is difficult. Because there are not intended to broadcast into space; broadcast antennas aim most of their transmitted power toward the surface. Most signal information is transmitted in bands on each side of the central frequency. What little detectable power reaches space is from many sources, not at the exact same frequencies, but in bands constrained by regulation by governments. Therefore, they are not coherent, so phase differences cause them to cancel each other out at great range.

Decades ago, it was thought that ETI would have an easier time spotting signals from over-the-horizon radars built during the Cold War, which directed much of their power into space. But those have since been superseded by frequency-hopping 'spread-spectrum' broadband radars that are undetectable by ETI.

Interplanetary radar used for asteroid searches are the present-day sources of highest power per solid angle. The radar emissions are not directed to stars at all, but to nearby asteroids. The METI 'messages' are directed at nearby stars. So, the possibility that the radars of the past can be detected by nearby stars is reduced by roughly the ratio of empty space to stellar cross-sections (habitable zones), a huge correction. Consequently, 'the probability of our (radar) getting into inhabited zones is insignificantly small' (Zaisev, 2007).

To quantify this, the Arecibo radar has a beam which intercepts $\sim 10^{-5}$ of the sky. The beam size at 100 ly is ~0.2 ly. The number of F through K stars within 100 ly is ~2,000 (Oliver and Billingham 1996, pg. 54). Therefore, these stars would be in 2,000 of the $10^5$ spots, a 2% fraction. However, the radar in on in its narrow-bandwidth 'unswitched' mode only one hour per year ($10^{-4}$ of a year) during setup of the 'switched' broader-band radar broadcasting (Billingham & Tarter 1992). An estimate of probability of intercept is $(2 \cdot 10^{-2})(10^{-4}) = 2 \cdot 10^{-6}$, or one in a half-million. However, this is an over-estimate because of two factors: 1) the radars are directed largely in the plane of the ecliptic, where the asteroids are, so most stars are not in the beams. 2) The beams are in their unswitched mode for only about a minute, so integration will be less effective (eq. 7). The conclusion is that the probability of intercept is much further reduced, so there is negligible chance of ETI noticing our asteroid search radars.

Recently others have agreed with the above, i.e., that our leakage radiation is not detectable by ETI: 'To assume that leakage automatically constitutes a ''reply from Earth'' to any SETI signal we might receive is unrealistic.' (Shostak 2011), as well as taking the opposite position: 'If there are any aliens within a few hundred light-years, these clues to our existence could be found with an antenna the size of Chicago.' (Shostak 2010).

Certainly, with ever-larger antenna area, at ever greater cost, advanced ETI can detect us. From the above analysis, we calculate that at 50-ly range, the antenna area must be ~1 km$^2$. To assess Shostak's claim, note that Chicago's area is 24,800 km$^2$ = $2.48 \cdot 10^{10}$ m$^2$. At the present value of SKA antennas, 2.4 k$/m$^2$, the cost is 60 T$, comparable to Earth's GNP of 70 T$. So if comparable to us, ETI would have to devote their entire science budget for a time perhaps of order a century to build Shostak's antenna, a sobering prospect.

The assertion of Loeb and Zaldarriaga (2006) that SKA can see leakage radiation at 100 pc (316 ly) is based on the assumption that the sources are continuous, so long integration times make the leakage detectable. However, this is not true of Earth leakage. Integrating over days to months doesn't work when the TV station you're observing is transmitting in your direction for a time typically ~hour, before it disappears around the limb of the Earth, as stated by Sullivan. Forgan and Nichol (2010) show that, even if Loeb and Zaldarriaga were right, the probability of detection is very low.

We should be mindful also of the future possibilities for increased leakage from Earth due to beaming power for space industrial purposes, such as power transfer. Examples are transferring energy from Earth-to-space, space-to-Earth, and space-to-space using high power microwave beams (Benford, 2008). Microwave beams have been studied for propelling spacecraft for launch to orbit, orbit-raising, and launch from orbit into interplanetary and interstellar space. The power levels are ~GW with high directivity, so that isotropic radiated power W~$10^{17}$ watts, would dwarf anything yet emitted into space. Observing such activities would appear to ETI as transient events.

**5.0 A Better Way Forward**

In conclusion, we have not announced ourselves either by leakage radiation or by intentional transmissions of recent years. We feel that conventional SETI, with powerful new search systems, is less risky and less expensive than METI.

We also feel that it is better to spend limited funds on listening, rather than transmitting. The advantages are:

- It's more expensive to transmit. For the same cost, much larger searches can be done, so detection is more likely.
- There is no waiting time for a response, as there is for transmitting.
- SETI searches for transients will yield good science astronomical data.
- Searching may allow more time for risk assessment.

We also advocate a moratorium on METI until an international consensus is reached.

## Acknowledgements

We thank Ian Morrison, Gregory Benford, Dominic Benford and Jill Tarter for helpful comments.

## References

Benford, G., Benford J. & Benford D. 2010 Searching for Cost Optimized Interstellar Beacons, *Astrobiology,* **10** 4, 490-498.

Benford, J., Benford G. & Benford D. 2010. Messaging with Cost Optimized Interstellar Beacons, *Astrobiology*, **10** 4, 475-490.

Benford J., 2008 Space Applications of High Power Microwaves, *IEEE Trans. on Plasma Sci.,* **36** 569-581.

Billingham, J. and Tarter, J. 1992 Detection of Earth with the SETI Microwave Observing System Assumed to be Operating Out in the Galaxy, *Acta Astronautica* **26** 185-188.

Billingham, J. and Heyns, R., et al. 1999 Social Implications of the Detection of an Extraterrestrial Civilization. Appendix, 129-132. *SETI Press* Mt. View, CA

Conley, C. 2010 NASA Planetary Protection Programs
*<http://www.planetaryprotection.nasa.gov.html>*http://www.planetaryprotection.nasa.html

DeBoer,D. 2010 personal communication).

Forgan D.H., Nichol R.C. 2010 A Failure of Serendipity: the Square Kilometre Array will struggle to eavesdrop on Human-like ETI, *IJA*, (2010), in press, also at arXiv:1007.0850v1

Hawking, S.W. 2010. Into the Universe. Part 1. Aliens. *Discovery TV Channel documentary. April 10*

Kraus, J. 1986, *Radio Astronomy, 2nd Edition*, Cygnus-Quasar, Durham, N.H.

Loeb A., Zaldarriaga M. 2007, *Journal of Cosmology and Astroparticle Physics*, 2007, 020

Oliver, B. and Billingham, J., Eds. 1996, *Project Cyclops,* NASA CR-114445.

Primary Mission Accomplished: 1969 - the Astronauts in Quarantine. <http://www.hq.nasa.gov.office/pao/History/SP-/4214><*http://www.hq.nasa.gov.office/pao/History/SP-/4214ch9-7.html*>http://www.hq.nasa.gov.office/pao/History/SP-/4214ch9-7.html

Schilizzi, R. et al. *Preliminary Specification of the SKA,* 2007, http://www.skatelescope.org/PDF/memos/100_Memo_Schilizzi.pdf

Shostak S. 2010 Don't Tell ET, http://www.huffingtonpost.com/seth-shostak/dont-tell-et_b_685774.html.

Shostak S. 2011 Limits to Interstellar Messages, *Acta Astronautica* **68** 366-371.

Sullivan W.T. III, Brown, S. & Wetherill C. 1978 Eavesdropping: The Radio Signature of Earth. *Science* 199, 377-388.

Tarter, J.C., et al. 2010. SETI turns 50: five decades of progress in the search for extraterrestrial intelligence, *SP*IE Optics and Photonics Symposium, San Diego. SPIE Conference Paper No.7819-1. In Press

Zaitsev, A. 2006, Report on Cosmic Call 1999, *http://www.cplire.ru/html/ra&sr/irm/report-1999.html*

Zaitsev, A. 2007 Sending and Searching for Interstellar Messages *Acta Astronautica* **63**, 614-617